\documentclass{juliacon}
\usepackage{amsmath}
\usepackage{float}
\begin{document}

\title{DisjunctiveProgramming.jl: Generalized Disjunctive Programming Models and Algorithms for JuMP}
\author{
  \large Hector D. Perez, Shivank Joshi, Ignacio E. Grossmann*
  \\\normalsize Carnegie Mellon University, Pittsburgh, PA, USA  
  \\\normalsize	*grossmann@cmu.edu
}

\maketitle

\section*{ABSTRACT}
We present a Julia package, \verb|DisjunctiveProgramming.jl|, that extends the functionality in \verb|JuMP.jl| to allow modeling problems via logical propositions and disjunctive constraints. Such models can then be reformulated into Mixed-Integer Programs (MIPs) that can be solved with the various MIP solvers supported by JuMP. To do so, logical propositions are converted to Conjunctive Normal Form (CNF) and reformulated into equivalent algebraic constraints. Disjunctions are reformulated into mixed-integer constraints via the reformulation technique specified by the user (Big-M or Hull reformulations). The package supports reformulations for disjunctions containing linear, quadratic, and nonlinear constraints.

\section*{Keywords}
JuMP, Mathematical Optimization, Generalized Disjunctive Programming

\section{Introduction}
The modeling of systems with discrete and continuous decisions is commonly done in algebraic form with mixed-integer programming (MIP) models. When the problems can be defined by purely linear constraints and a linear objective function, they are referred to as mixed-integer linear programs (MILP). When nonlinearities arrise in either the feasible space or the objective function, they are called mixed-integer nonlinear programs (MINLP).
\vskip 6pt
A more systematic approach to modeling such systems is to use Generalized Disjunctive Programming (GDP) \cite{chen_grossmann_2019, grossmann_trespalacios_2013}, which generalizes the Disjunctive Programming paradigm proposed by Balas \cite{balas_2018}. GDP enables the modeling of systems from a logic-based level of abstraction that captures the fundamental rules governing such systems via algebraic constraints and logic. This formulation is useful for expressing problems in an intuitive way that relies on their logical underpinnings without needing to introduce mixed-integer constraints. GDP models are often easier to understand as related constraints are grouped into disjuncts that describe clearly defined subsets of the feasible space. The models obtained via GDP can be reformulated into the pure algebraic form best suited for the application of interest.  It is also often possible to exploit the explicit logical structure of a GDP model to provide tighter relaxations than corresponding MIP models, which may improve convergence speed and robustness for solutions via advanced solution algorithms \cite{chen_grossmann_2019}.
\vskip 6pt
Within the optimization community, there is a high volume of ongoing research that relies on GDP to formulate models for a variety of applications. Due to the combinatorial nature of system design problems, the GDP paradigm has been applied to the synthesis of complex processes and networks \cite{MATOVU2022107856, ZHOU202269}, the planning and optimal control of energy systems \cite{CHO2022841, kim2022generalized}, and the modeling of chemical synthesis under uncertainty \cite{CHEN2022107616}. These and numerous other applications of GDP illustrate the benefit of having a robust package for GDP that removes much of the overhead associated with developing and testing GDP models. Although packages with GDP capabilities exist for \verb|Pyomo| \cite{chen2022pyomo} and \verb|GAMS| \cite{vecchietti1999logmip}, having such a package available in Julia can greatly accelerate research in optimization, where packages like \verb|JuMP.jl| \cite{dunning_huchette_lubin_2017} are gaining significant traction.
\vskip 6pt
This paper provides background on the GDP paradigm, and the techniques for reformulating and solving such models. It then presents the package \verb|DisjunctiveProgramming.jl| as an extension to \verb|JuMP.jl| for creating models for optimization that follow the GDP modeling paradigm and can be solved using the vast list of supported solvers \cite{dunning_huchette_lubin_2017}. A case study demonstrates the use of the package for chemical process superstructure optimization.

\section{Generalized Disjunctive Programming}
The GDP form of modeling is an abstraction that uses both algebraic and logical constraints to capture the fundamental rules governing a system. The two main reformulation strategies to transform GDP models into their equivalent MIP models are the Big-M reformulation \cite{nemhauser_1999, TRESPALACIOS201598} and the Hull reformulation \cite{LEE20002125}, the latter of which yields tighter models at the expense of larger model sizes \cite{grossmann_lee_2003}. 
\vskip 6pt

\subsection{Model}

The general notation for a GDP problem is given in Eq. \eqref{eq:general_gdp} - \eqref{eq:general_gdp1}.
\begin{align}
    \label{eq:general_gdp}
    \min \ &f(x) \\
    \text{s.t.} \ &g(x) \leq 0 \\
    &\bigvee_{i \in J_k}
    \begin{bmatrix}
        Y_{ik} \\
        h_{ik}(x) \leq 0
    \end{bmatrix} \quad \forall k \in K \\
    & \Omega(Y) = true \\
    & Y_{ik} \in \{true, false\} \quad \forall i \in J_k, k \in K\\
    \label{eq:general_gdp1}
    & x \in X \subseteq \mathbb{R}^n
\end{align}
Here $f(x)$ is the objective function to be minimized over the continuous variable $x$, $g(x)$ represents the global constraints, $h(x)$ represents the disjunct-specific constraints, and $Y$ is the Boolean variable governing each disjunction. In this notation, there are $k$ disjunctions with $i$ disjuncts in each. Constraints $h_{ik}(x) \le 0$, are applied only if the Boolean indicator variable for the respective disjunct, $Y_{ik}$, is denoted as being active (i.e., \textit{true}) \cite{chen_grossmann_2019}. The set of logical constraints, $\Omega(Y)$, describe the logical relationships between the selections of indicator variables. These can take the form of logical propositions or constraint programming expressions.
\vskip 6pt

In the case of a linear objective function and constraint set, the GDP model can be written as Eq. \eqref{eq:lgdp} - \eqref{eq:lgdp2}.
\begin{align}
    \label{eq:lgdp}
    \min \ & c^Tx \\
    \text{s.t.} \ &Ax \leq b \\
    &\bigvee_{i \in J_k}
    \begin{bmatrix}
        Y_{ik} \\
       B_{ik}x \leq d_{ik}
    \end{bmatrix}, \quad \forall k \in K \\
    & \Omega(Y) = true \\
    \label{eq:lgdp2}
    & Y_{ik} \in \{true, false\} \quad \forall i \in J_k, k \in K
\end{align}
\vskip 6pt

\subsection{Linear GDP reformulation example}
The simplest example of a linear GDP system is given in \eqref{eq:ex} - \eqref{eq:y}, where $Y_i$ is a Boolean indicator variable that enforces the constraints in the disjunct ($Ax \le b$ or $Cx \le d$) when $true$.
\vskip 6pt
\begin{equation}
    \label{eq:ex}
    \begin{bmatrix}
    Y_1 \\ Ax \leq b
    \end{bmatrix}
    \lor
    \begin{bmatrix}
        Y_2 \\ Cx \leq d
    \end{bmatrix}
\end{equation}
\begin{equation}
    \label{eq:x}
    0 \leq x \leq U
\end{equation}
\begin{equation}
    \label{eq:simple_xor}
    Y_1 \ \underline{\vee} \ Y_2
\end{equation}
\begin{equation}
    \label{eq:y}
    Y_1, Y_2 \in \{true, false\}
\end{equation}
\vskip 6pt

\begin{figure}
    \centering
    \includegraphics[scale=0.5]{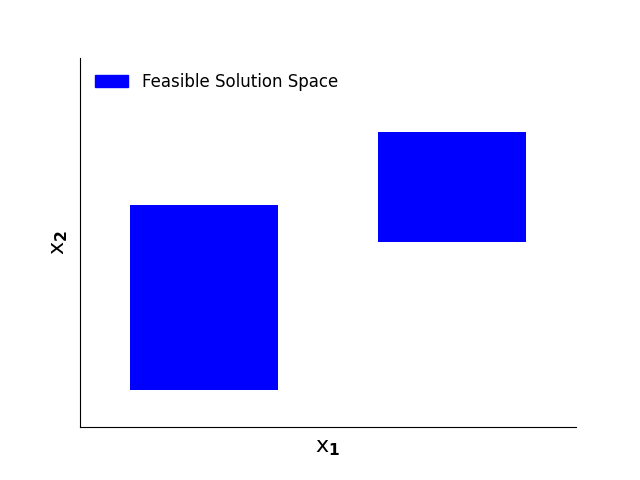}
    \caption{Feasible solution space for example disjunction}
    \label{fig:reform_figure}
\end{figure}

For visualization purposes and without loss of generality, the simple linear example is used to illustrate the Big-M and Hull reformulations. Figure \ref{fig:reform_figure} illustrates the feasible space of a simple linear GDP with one disjunction and two continuous variables, $x_1$ and $x_2$. The rectangle on the left is described by the constraints in the left disjunct, $Ax \leq b$. The rectangle on the right is defined by the constraints in the right disjunct, $Cx \le d$. The non-overlapping nature of the two regions is supported by the exclusive-OR relationship in Eq. \eqref{eq:simple_xor}.

\vskip 6pt

 \subsubsection{Big-M Reformulation}
\begin{figure}
    \centering
    \includegraphics[scale=0.5]{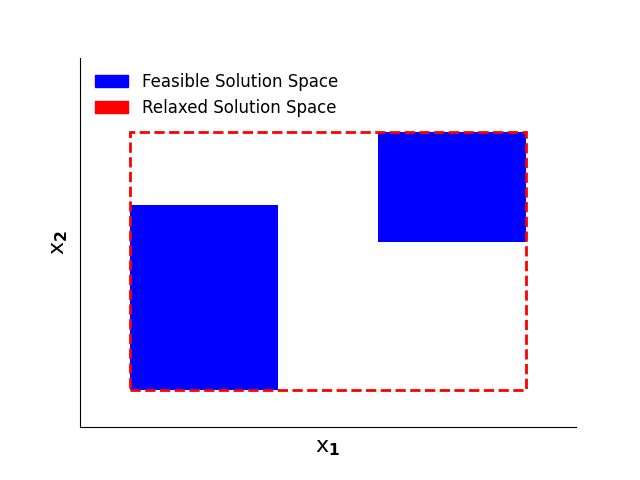}
    \caption{Relaxed solution space using Big-M Reformulation}
    \label{fig:bigm}
\end{figure}

 The Big-M reformulation for this problem is given by \eqref{eq:x}, \eqref{eq:ex_bigm1} - \eqref{eq:ex_bigm4}, where $M$ is a sufficiently large scalar that makes the particular constraint redundant when its indicator variable is not selected (i.e., $y_i = 0$). Note that the Boolean variables, $Y_i$, are replaced by binary variables, $y_i$. When the integrality constraint in Eq. \eqref{eq:ex_bigm4} is relaxed to $0 \leq x_1, x_2 \leq 1$, the resulting feasible region can be visualized by projecting the relaxed model onto the $x_1, x_2$ plane. This results in the region encapsulated by the dashed line in Figure \ref{fig:bigm}. It should be noted that the relaxed feasible region is not as tight as possible around the original feasible solution space. The choice of the large $M$ value determines the tightness of this relaxation, and the minimal value of $M$ for the optimal relaxation can be found through interval arithmetic when the model is linear. For nonlinear models, the tightest $M$ can be obtained by solving the maximization problem $\{\max h_{ik}(x): x \in X\}$. An alternate method for tight Big-M relaxations is given in \cite{TRESPALACIOS201598}.
 
\begin{equation}
    \label{eq:ex_bigm1}
    Ax \leq b + M \cdot (1 - y_1)
\end{equation}
\begin{equation}
    \label{eq:ex_bigm2}
    Cx \leq d + M \cdot (1 - y_2)
\end{equation}
\begin{equation}
    \label{eq:ex_bigm3}
    y_1 + y_2 = 1
\end{equation}
\begin{equation}
    \label{eq:ex_bigm4}
    y_1, y_2 \in \{0,1\}
\end{equation}

\subsubsection{Hull Reformulation}
\begin{figure}
    \centering
    \includegraphics[scale=0.5]{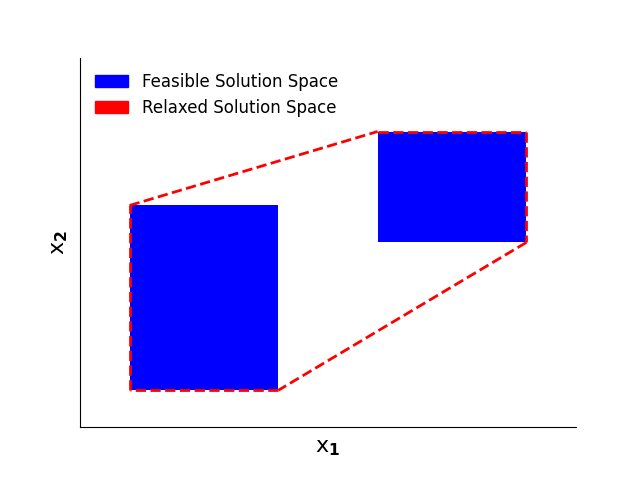}
    \caption{Relaxed solution space using Hull Reformulation}
    \label{fig:chr}
\end{figure}

The Hull reformulation is given by \eqref{eq:x}, \eqref{eq:ex_bigm3} - \eqref{eq:ex_hull3}, which requires lifting the model to a higher-dimensional space. When projected to the original space, the continuous relaxation of the model is tighter than its Big-M equivalent \cite{grossmann_trespalacios_2013}. The reformulation relaxation can be visualized by the region encapsulated by the dashed line in Figure \ref{fig:chr}. Note that this reformulation provides a tighter relaxation than the Big-M reformulation in Figure \ref{fig:bigm}. Also note that describing the geometry of this relaxation is more complex than the Big-M relaxation, which is made possible by the increased number of constraints and variables in the model.

\begin{equation}
    \label{eq:ex_hull1}
    Ax_1 \leq by_1
\end{equation}
\begin{equation}
    \label{eq:ex_hull0}
    Cx_2 \leq dy_2
\end{equation}
\begin{equation}
    \label{eq:ex_hull2}
    x = x_1 + x_2
\end{equation}
\begin{equation}
    \label{eq:ex_hull3}
    0 \leq x_i \leq U y_i \quad \forall i \in \{1,2\}
\end{equation}

\vskip 6pt

\subsection{Logic constraint reformulation}

\subsubsection{Propositional Logic}
The logic propositions within the set of decision selection relationships, $\Omega$, must be converted into conjunctive normal form (CNF) to enable reformulating a GDP model as a MIP model. This means that each clause within the set of propositions must be formulated into a conjunction of disjunctions. This process can be accomplished by following the simplifying rules of propositional logic (De Morgan's laws). For boolean variables $A$, $B$, and $C$ the following rules are used for converting to CNF (in the order given below),
\begin{align*}
    A \leftrightarrow B & \text{ is replaced by } (A \rightarrow B) \land (B \rightarrow A) \\
    A \rightarrow B & \text{ is replaced by } \lnot A \lor B \\
    \lnot(A \lor B) & \text{ is replaced by } \lnot A \land \lnot B \\
    \lnot(A \land B) & \text{ is replaced by } \lnot A \lor \lnot B \\
    (A \land B) \lor C & \text{ is replaced by } (A \lor C) \land (B \lor C)
\end{align*}

Once the logic propositions are converted to CNF, each clause can be converted into an algebraic constraint with the following equivalence (Note: any negated Boolean variables, $\neg Y_i$, are replaced with $1-y_i$ in the reformulation),

\begin{align*}
    \bigvee_{i \in I} Y_i & \ \ \text{becomes} \ \ \sum_{i\in I} y_i \geq 1 \\
\end{align*}

Alternate approaches exist for converting propositional logic statements into CNF, which involve preserving clause satisfiability rather than clause equivalence. These approaches prevent exponential size increase in clauses and yield logically consistent results \cite{jackson_sheridan_2005}.

\subsubsection{Constraint Programming}
Selection constraints analogous to those used in Constraint Programming (CP) can also be included in $\Omega(Y)$. These constraints are of the form "allow exactly $n$ elements in a list of Boolean variables to be $true$." This type of constraint overcomes the limitations of the Boolean exclusive-OR ($\underline{\vee}$) operator, which can only enforce that an odd number of elements in a list of Booleans be $true$. Other CP-like constraints can be obtained by replacing "exactly" with "at most" or "at least". These constraints are reformulated as follows,

\begin{align*}
    \text{exactly}(n, Y) & \ \ \text{becomes} \ \ n = \sum_i Y_i \\
    \text{atleast}(n, Y) &  \ \ \text{becomes} \ \ n \leq \sum_i Y_i \\
    \text{atmost}(n, Y) &  \ \ \text{becomes} \ \ n \geq \sum_i Y_i
\end{align*}

Exclusive-OR constraints as the one given in Eq. \eqref{eq:simple_xor} are more generally modeled as $exactly(1,\{Y_1,Y_2\})$.

\subsection{Solution Techniques}

\subsubsection{Disjunctive branch and bound}
The disjunctive branch and bound method closely mirrors the standard branch and bound approach for the solution of mixed-integer programming problems \cite{grossmann_lee_2003}. A search tree is initialized by solving the continuous relaxation of the Big-M or Hull reformulation of the original GDP to obtain a lower bound on the optimum. Branching is then done on the disjunction with an indicator binary variable closest to 1. Two nodes are created at this point: one where the respective indicator Boolean variable is fixed to $true$ (the disjunct is enforced) and another where it is fixed to $false$ (the disjunct is removed from the disjunction). Each node is reformulated and solved to obtain a candidate lower bound. If the solution to a node results in a feasible solution that satisfies all integrality constraints, the solution is an upper bound on the optimum. Any non-integral solutions that exceed an upper bound are pruned from the search tree. The process is repeated until the lower and upper bounds are within the desired tolerance.

\subsubsection{Logic-based outer approximation}
Logic-based outer approximation is another algorithm which mirrors a standard technique for solving mixed-integer nonlinear programming problems \cite{E.Grossmann2009}. This approach starts by identifying a set of reduced Non-Linear Programming (NLP) sub-problems obtained by fixing Boolean variables in the different disjunctions such that each disjunct is selected at least once across the set of sub-problems (set covering step). 
Each sub-problem is solved to obtain an upper bound and a feasible point, about which the objective and constraints of the original GDP are linearized, and solve the resulting problem (via direct reformulation to MILP or via disjunctive branch and bound) to find a lower bound. If the lower and upper bound solutions have not converged, the Boolean variables from the previous solution are fixed and the resulting NLP is solved to find a potentially tighter upper bound solution. The procedure is repeated until convergence is obtained.

\subsubsection{Hybrid cutting planes}
The cutting planes method is an algorithm for tightening the relaxed solution space of a problem reformulated with Big-M before solving it by adding additional constraints which remove parts of the relaxed space that are disjoint from the actual feasible solution space. These "cuts" to the relaxed solution space are derived from the tighter, hull relaxation of the problem. This algorithm provides a middle-ground for the tradeoff between the complexity and corresponding computational expense of the Hull reformulation with the less tight Big-M reformulation. \cite{trespalacios_grossmann_2016}. 

\section{DisjunctiveProgramming.jl}
The following section describes the features of the \verb|DisjunctiveProgramming.jl| package and illustrates its syntax with an example from the chemical processing industry for superstructure optimization. The use of nested disjunctions is also shown.

\subsection{Features}
\verb|DisjunctiveProgramming.jl| allows for defining JuMP models with disjunctions that are directly reformulated via Big-M or Hull methods via the \verb|@disjunction| macro or \verb|add_disjunction!| function. Big-M values can be specified either for the entire disjunction, for each disjunct, or for each constraint in each disjunct. Alternately, if the constraints are linear, the code can use the variable bounds to perform interval arithmetic on each constraint to determine the tightest possible $M$ value to use \cite{agarwal2010automating}. For nonlinear GDP constraints, the epsilon approximation formulation for the perspective function in the Hull reformulation is used \cite{furman_sawaya_grossmann_2020}. Users can specify an epsilon tolerance value to use. Perspective functions are generated by relying on manipulation of symbolic expressions via \verb|Symbolics.jl| \cite{10.1145/3511528.3511535}.

\vskip 6pt
Logical propositions can be added to JuMP models using expressions involving the disjunction indicator variables and the standard Boolean operators ($\Leftrightarrow, \Rightarrow, \vee, \wedge, \text{ and } \neg$) in an \verb|@proposition| macro. These are automatically converted into CNF and added as integer algebraic constraints to the model. The constraint programming constraints can also be added using the \verb|choose!| function. The expressions are also automatically reformulated into integer algebraic constraints.

\vskip 6pt
Nesting of disjunctions is also supported.

\subsection{Example}
\begin{figure}
    \centering
    \includegraphics[scale=0.4]{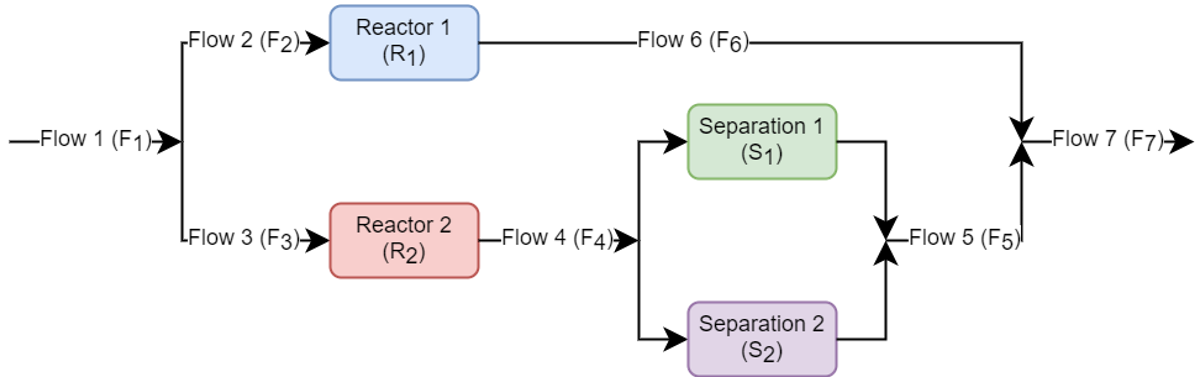}
    \caption{Illustrative superstructure optimization problem}
    \label{fig:superstruct_opt_diagram}
\end{figure}

To illustrate the syntax in \verb|DisjunctiveProgramming.jl| (Version 0.3.3), consider the simple superstructure optimization problem for the chemical process given in Figure \ref{fig:superstruct_opt_diagram}. In this problem a chemical plant with two candidate reactor technologies ($R_1$ and $R_2$) must be designed. If the second reactor technology is chosen, a separation system must also be installed, for which two separation technologies ($S_1$ and $S_2$) are available. The GDP model seeks to maximize the product flow ($F_7$), while discounting for reactor ($C_R$) and separator ($C_S$) installation costs as given in \eqref{eq:example_obj}, subject to the nested disjunction in \eqref{eq:example_gdp} and the global mass balances in \eqref{eq:example_global}
- \eqref{eq:example_global1}. The system variables are the flows on each stream $i$ ($F_i$) and the installation costs, with their respective bounds given in \eqref{eq:example_var1} - \eqref{eq:example_var3}. The fixed cost and process yield parameters are given by $\gamma$ and $\beta$, respectively.

\begin{equation}
    \label{eq:example_obj}
    \max F_7 - C_R - C_S
\end{equation}
\begin{equation}
    \label{eq:example_gdp}
    \begin{bmatrix}
        Y_{R1} \\
        F_6 = \beta_{R1} F_2 \\
        F_3 = 0 \\
        F_4 = 0 \\
        F_5 = 0 \\
        C_R = \gamma_{R1} \\
        CS = 0
    \end{bmatrix} \lor
    \begin{bmatrix}
        Y_{R2} \\
        F_2 = 0 \\
        F_6 = 0 \\
        F_4 = \beta_{R2} F_3 \\
        C_R = \gamma_{R2} \\
        \begin{bmatrix}
            Y_{S1} \\
            F5 = \beta_{S1} F_4 \\
            C_S = \gamma_{S1}
        \end{bmatrix} \lor
        \begin{bmatrix}
            Y_{S2} \\
            F5 = \beta_{S2} F_4 \\
            C_S = \gamma_{S2}
        \end{bmatrix}
    \end{bmatrix}
\end{equation}
\begin{equation}
    \label{eq:example_global}
    F_1 = F_2 + F_3
\end{equation}
\begin{equation}
    \label{eq:example_global1}
    F_7 = F_5 + F_6
\end{equation}
\begin{equation}
    \label{eq:example_var1}
    0 \leq F_i \leq 10 \quad \forall i \in \{1,...,7\}
\end{equation}
\begin{equation}
    \label{eq:example_var2}
    0 \leq C_S \leq C_S^{max}
\end{equation}
\begin{equation}
    \label{eq:example_var3}
    C_R^{min} \leq C_R \leq C_R^{max}
\end{equation}
\vskip 6pt

The above system can be modeled and reformulated via the Big-M reformulation using \verb|DisjunctiveProgramming.jl|. The resulting JuMP model is then optimized using the HiGHS open-source MILP solver \cite{huangfu2018parallelizing} as shown below.

\begin{enumerate}
    \item Create the JuMP model and define the model variables and global constraints (mass balances).
    
\begin{lstlisting}[language = Julia]
using DisjunctiveProgramming, JuMP, HiGHS

# create model
m = JuMP.Model(HiGHS.Optimizer)
# add variables to model
@variable(m, 0 <= F[i = 1:7] <= 10)
@variable(m, 0 <= CS <= CSmax)
@variable(m, CRmin <= CR <= CRmax)

# add constraints to model
@constraints(m,
    begin
        F[1] == F[2] + F[3]
        F[7] == F[5] + F[6]
    end
)
\end{lstlisting}
    \item Define the inner (nested) disjunction for the separation technologies in the superstructure using the \verb|@disjunction| macro.   
\begin{lstlisting}[language = Julia]
@disjunction(m,
    begin
        F[5] == β[:S1]*F[4]
        CS == γ[:S1]
    end,
    begin
        F[5] == β[:S2]*F[4]
        CS == γ[:S2]
    end,
    reformulation = :big_m, # reformulation type
    name = :YS # symbol for indicator variable
)
\end{lstlisting}
    \item Define constraints in the outer disjunctions.
\begin{lstlisting}[language = Julia]
# define constraints in left disjunct
R1_con = @constraints(m,
    begin
        F[6] == β[:R1]*F[2]
        [i = 3:5], F[i] == 0
        CR == γ[:R1]
        CS == 0
    end
)

# define constraints in right disjunct
R2_con = @constraints(m,
    begin
        F[6] == β[:R2]*F[3]
        CR == γ[:R2]      
    end
)
\end{lstlisting}
    \item Build the main disjunction using the constraint blocks defined in (3) and the \verb|add_disjunction!| function. Note that the reformulated constraints for the nested disjunction are stored in the \verb|.ext| dictionary of the model under the name of the disjunction (\verb|:YS| in this case).
\begin{lstlisting}[language = Julia]
add_disjunction!(m,
    R1_con,
    (
        R2_con, #general constraints in R2 disj.
        m.ext[:YS] #reformulated inner disj.
    ),
    reformulation = :big_m, # reformulation type
    name = :YR # symbol for indicator variable
)
\end{lstlisting}
    \item Add the selection logical constraints using the \verb|choose!| function. The first constraint enforces that only one reactor is selected (i.e., $Y_{R_1} \ \underline{\vee} \ Y_{R_2}$). The second constraint enforces that the separation system be defined only if the second reactor ($R_2$) is selected. This constraint is equivalent to the proposition $Y_{R_2} \Leftrightarrow Y_{S_1} \ \underline{\vee} \ Y_{S_2}$.
\begin{lstlisting}[language = Julia]
choose!(1, YR[1], YR[2]; mode = :exactly)
choose!(YR[2], YS[1], YS[2]; mode = :exactly)
\end{lstlisting}
    \item Add the objective function and optimize. 
\begin{lstlisting}[language = Julia]
@objective(m, Max, F[7] - CS - CR)
optimize!(m)
\end{lstlisting}
\end{enumerate}

\section{Future Work}
The next steps for the \verb|DisjunctivePrograming.jl| package rely on extending \verb|JuMP.jl| further to allow creating GDP models that are not necessarily reformulated at model creation. Such models will allow using the different GDP solution strategies, such as direct reformulation to MI(N)LP, disjunctive branch and bound, logic-based outer approximation, hybrid cutting planes, and basic steps. The updated package will leverage existing JuMP extension infrastructure and make it possible to define indexing notation for disjunctions, a new Boolean variable type, and a new disjunciton constraint type. These improvements are expected to make the package more usable, flexible, and performant for advanced applications of GDP in JuMP.

\section{Related Work}
The popular Python package \verb|Pyomo| \cite{bynum2021pyomo, hart2011pyomo} is widely used for optimization development and includes an extension for generalized disjunctive programming \cite{chen2022pyomo}. \verb|GAMS| \cite{Bussieck2004} is a widely used optimization modeling language with support for GDP under the \verb|GAMS EMP| solver that uses \verb|LogMIP| \cite{vecchietti1999logmip}. Research is also being conducted to integrate modern process simulation technology, such as \verb|Aspen|, within the GDP paradigm \cite{NAVARROAMOROS201413}.

\section{Conclusion}
\verb|DisjunctiveProgramming.jl| is an extension to \verb|JuMP| for creating models for optimization that are formulated according to the generalized disjunctive programming paradigm. The package provides several options for reformulations including the Big-M and Hull relaxations. This package can be used to model problems, reformulate them, and optimize them using existing mathematical programming infrastructure in \verb|JuMP|. This can be useful for industrial and academic applications of GDP, such as superstructure optimization. The capabilities of this package allow for this modeling paradigm to be exploited using \verb|Julia|'s efficient dynamically-typed systems for rapid development, building, and testing of optimization models.


\bibliographystyle{juliacon}
\bibliography{ref.bib}

\end{document}